\documentclass[a4paper]{AO4ELT}  
\usepackage{microtype}
\usepackage{biblatex}
\usepackage{graphicx}
\usepackage{pst-all} 
\usepackage{amsmath,amsfonts,amssymb}
\usepackage{graphicx}
\usepackage[colorlinks=true, allcolors=blue]{hyperref}
\usepackage{lineno}
\usepackage{siunitx}
\usepackage{subcaption}
\usepackage[colorlinks=true, allcolors=blue]{hyperref}
\makeatletter         
\def\@maketitle{
\includegraphics[width = 165mm]{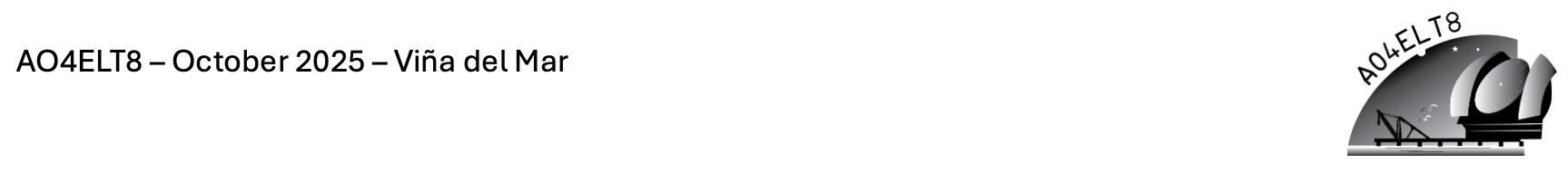}\\[8ex]
\begin{center}
{\Huge \bfseries \sffamily \@title }\\[4ex] 
{\Large  \@author}\\[4ex] 
\@date
\end{center}}

\title{Progress report on the integration of the IRTF adaptive secondary mirror}

\author[a]{Ellen Lee}
\author[a]{Mark Chun}
\author[a]{Ruihan Zhang}
\author[b]{Olivier Lai}
\author[a]{Michael Connelley}
\author[a]{Tony Denault}
\author[c]{Jacob Taylor}
\author[a]{John Rayner}
\author[d]{Max Baeten}
\author[d]{Arjo Bos}
\author[d]{Matias Kidron}
\author[d]{Fred Kamphues}
\author[d]{Stefan Kuiper}
\author[d]{Wouter Jonker}
\author[a]{Alan Ryan}
\author[e]{Philip Hinz}
\affil[a]{University of Hawai'i at Manoa - Institute for Astronomy, Honolulu, HI, USA}
\affil[b]{Observatoire de la Côte d'Azur, Nice, France}
\affil[c]{W. M. Keck Observatory, Waimea, HI, USA}
\affil[d]{TNO, Delft, Netherlands}
\affil[e]{University of California - Santa Cruz, Santa Cruz, CA, USA}

\authorinfo{Further author information: (Send correspondence to E.L.)\\E.L.: E-mail: ellenlee@hawaii.edu\\M.C.: E-mail: markchun@hawaii.edu}

\begin{document} 
\maketitle

\begin{abstract}
IRTF-ASM-1 is the first on-sky adaptive secondary mirror using the hybrid variable reluctance (HVR) actuators developed by the Netherlands Organization for Applied Scientific Research (TNO). Since its first light in April 2024, the ASM has continued to work consistently well with no hardware issues. The primary purpose of IRTF-ASM-1 is to serve as a demonstration of the HVR actuator technology, both in terms of verifying its robustness and testing calibration methods that are relevant to larger ASMs. However, as the ASM has proven to be easy to handle and reliable in its performance, we are moving toward integrating the ASM for long-term use at IRTF. We present closed loop results with the ASM and IRTF's off-axis facility wavefront sensor FELIX. Correcting the first seven Zernike modes through coma at 90-\qty{180}{\hertz}, we are able to enhance the seeing by a factor of 1.8 in FWHM under \qty{0.5}{\arcsecond} seeing conditions. We also performed the first science observations with the ASM in ``static" mode and demonstrated that we can improve the throughput of IRTF's slit spectrograph by approximately a factor of 2, although this requires good reference slopes in FELIX. In the near future, we plan to optimize the calibration of reference slopes in FELIX and streamline the software so that the system can be operated by a non-AO expert.
\end{abstract}


\section{INTRODUCTION}
\label{sec:intro}  
Adaptive secondary mirrors (ASMs) are highly advantageous for their ability to feed improved image quality to every instrument on the telescope without loss of throughput. ASMs are particularly well suited to provide correction of wide fields of view because they are optically conjugate to the ground, for example, through ground layer adaptive optics (GLAO) which can improve the seeing by a factor of two on Maunakea \cite{chun2018sky}. As a result, ASMs are becoming increasingly popular with several large observatories planning to install them in the near future, including the Subaru Telescope \cite{oya2024ultimate} and W. M. Keck Observatory \cite{hinz2024keck}. 

IRTF-ASM-1 is a prototype adaptive secondary mirror designed for the NASA Infrared Telescope Facility (IRTF), a 3.2-meter infrared-optimized telescope on Maunakea. The ASM uses 36 of TNO's HVR actuators with a \qty{39}{\milli\meter} spacing across three rings. The \qty{3.3}{\milli\meter} thick borosilicate shell is \qty{244}{\milli\meter} in diameter. This ASM was developed as the first demonstration of the HVR actuator technology in an observatory. The actuators are advantageous for their large linear range of stroke--at least $\pm\qty{5}{\micro\meter}$ between actuators and up to $\pm\qty{15}{\micro\meter}$ across the entire facesheet--which permits a simple and robust ASM design. In April 2024, we successfully closed loop the loop with the ASM on-sky using a 12$\times$12 Shack-Hartmann wavefront sensor (SHWFS) and achieved diffraction-limited image quality at H-band over \qty{0.5}{\arcsecond} to \qty{1}{\arcsecond} seeing conditions \cite{lee2024first}. Since then, we have had six separate observing runs with the ASM. The day crew installs the ASM onto a spare top end ring before each run; the ASM is also brought down to Hilo for lab testing during the intervening months. Our observing schedule is provided in Table~\ref{tab:schedule}. The performance of the ASM has remained consistent since we first received it in February 2024 and we have not experienced any problems with the hardware.

\begin{table}[h]
\caption{Timeline of engineering with IRTF-ASM-1. In total, we have had six separate runs totaling 21 nights on the telescope of which 4 have been completely lost to weather.} 
\label{tab:schedule}
    \begin{center}       
        \begin{tabular}{|r|p{0.75\linewidth}|} 
        \hline
        \rule[-1ex]{0pt}{3.5ex}  Apr 2024 & First light. Closed loop with 12$\times$12 WFS.  \\
        \hline
        \rule[-1ex]{0pt}{3.5ex}  May 2024 & ($\times$2 runs) Loop tuning and testing modal control/piston filtering. Static ASM tests. \\
        \hline
        \rule[-1ex]{0pt}{3.5ex}  May 2025 & Closed loop with FELIX active optics using facility instrument software. Testing on-sky calibration methods for more actuators. \\
        \hline
        \rule[-1ex]{0pt}{3.5ex}  Jul 2025 & Further testing of interaction matrices and chopping the ASM.  \\
        \hline
        \rule[-1ex]{0pt}{3.5ex}  Oct 2025 & Closed loop with FELIX at \qty{200}{\hertz} using pyRTC.  \\
        \hline 
        \end{tabular}
        \end{center}
\end{table}

Because IRTF-ASM-1 has been robust and is easily handled, we are interested in installing it long-term at IRTF to improve the performance of the facility. IRTF does not have existing AO capability and as such its instruments are designed for seeing-limited observations. IRTF's facility instruments can still benefit from the improved image quality provided by the ASM. First, the telescope suffers from significant static aberrations which can be corrected by performing active optics correction with the ASM. Ref.~\cite{dinh2024measuring} shows that the full-width at half-maximum (FWHM) of the point spread function (PSF) can degrade by over a factor of two when pointing the telescope \qty{60}{\degree} from zenith. Additionally, if the ASM is operated more quickly ($\sim$100-200 Hz), it can provide enhanced seeing through the correction of wind shake and turbulence in the ground layer.

In this paper, we report our current progress in integrating IRTF-ASM-1 with FELIX, a low-order facility WFS that was originally designed to slowly monitor static aberrations. In Sec.~\ref{sec:overview}, we provide a description FELIX and how it is used to operate the ASM. We then describe the performance of the system in Sec.~\ref{sec:performance} and its ability to enhance the throughput of IRTF's science instruments in Sec.~\ref{sec:science}. Finally, Sec.~\ref{sec:challenges} discusses challenges with using the ASM at IRTF along with potential solutions.

\section{System overview} \label{sec:overview}

\begin{figure}[h]
    \centering
    \includegraphics[width=0.65\linewidth]{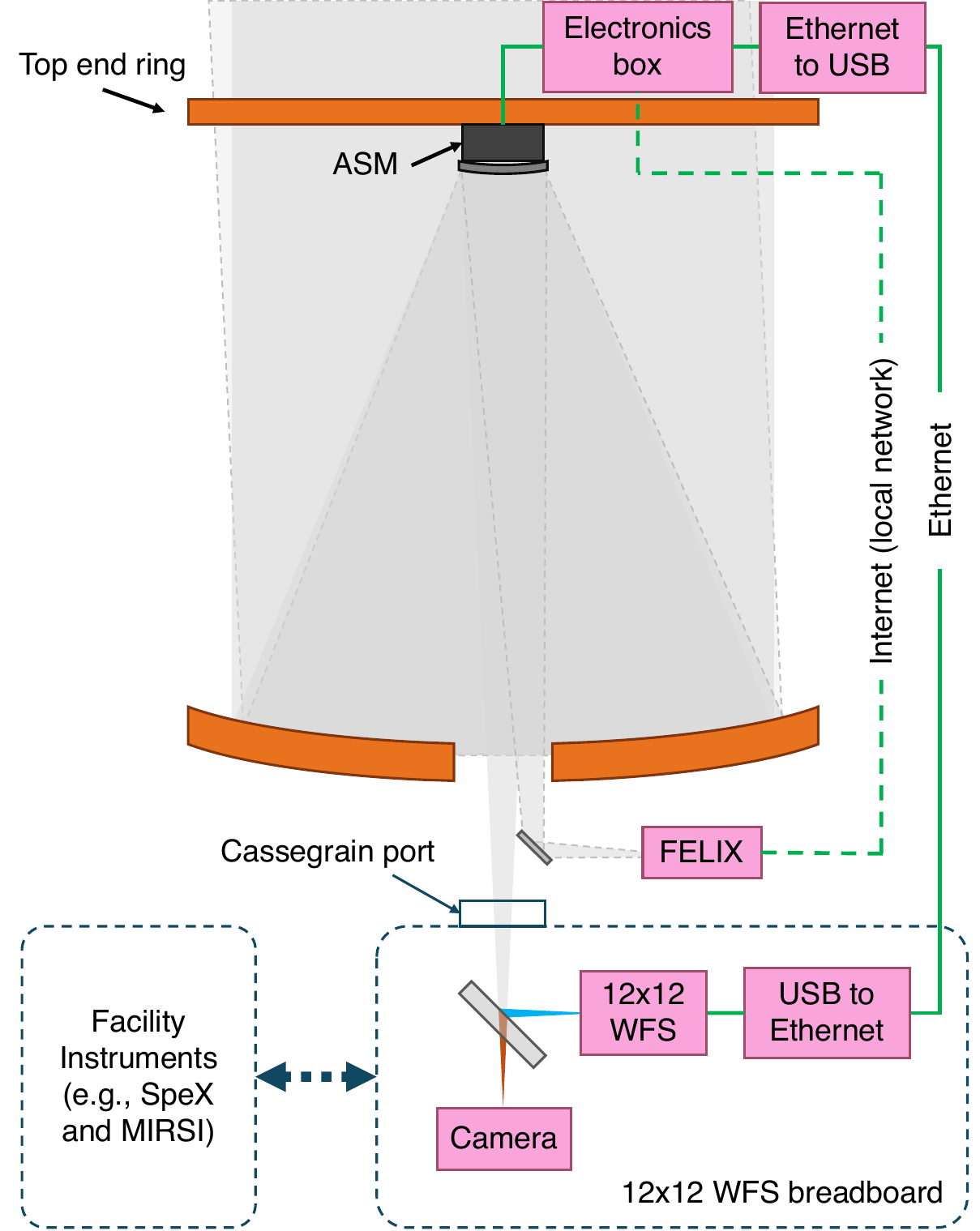}
    \vspace*{2mm}
    \caption{A diagram of the system including the ASM, a 12$\times$12 WFS and infrared scoring camera, and the off-axis facility WFS FELIX. FELIX's computer is located next to FELIX on the telescope. Because the ASM's electronics are connected to the 12$\times$12 RTC, FELIX's RTC currently communicates with the ASM via internet. This interface will likely be changed in the near future. The $12\times12$ wavefront sensor package is used at the Cassegrain focus of the telescope where all other science instruments are located. Using a facility instrument with the ASM requires us to swap the 12$\times$12 WFS package out, rendering it unavailable for use. FELIX can always be used simultaneously with facility instruments as it is off-axis.}
    \label{fig:system-diagram}
\end{figure}

Figure~\ref{fig:system-diagram} shows a diagram of the system on the telescope. A 12$\times$12 SHWFS is installed at the Cassegrain focus of the telescope as are the rest of IRTF's instruments. An infrared ($\lambda=$\qty{1.65}{\micro\meter}) scoring camera is also included on the 12$\times$12 WFS's breadboard to evaluate image quality from closing the loop with this WFS or with FELIX, an off-axis 2$\times$2 SHWFS. The facility instruments are simultaneously mounted on a mechanism at the bottom of the telescope that allows them to be manually swapped in and out in a process that takes half an hour. The 12$\times$12 WFS exists to monitor the performance of the ASM and is only temporarily at IRTF.

The ASM is mounted on a spare top end ring with a box containing the electronics for the ASM along with the real-time controller (RTC) for the $12\times12$ WFS. Installing the ASM currently requires a top end exchange because it does not fit on the hexapod that is used to collimate the telescope. A new hexapod will be ordered soon to accommodate the ASM allowing it to remain installed on the telescope for nightly use. 

\begin{figure}[h!]
    \centering
    \begin{subfigure}{\textwidth}
        \centering
        \includegraphics[width=0.7\linewidth]{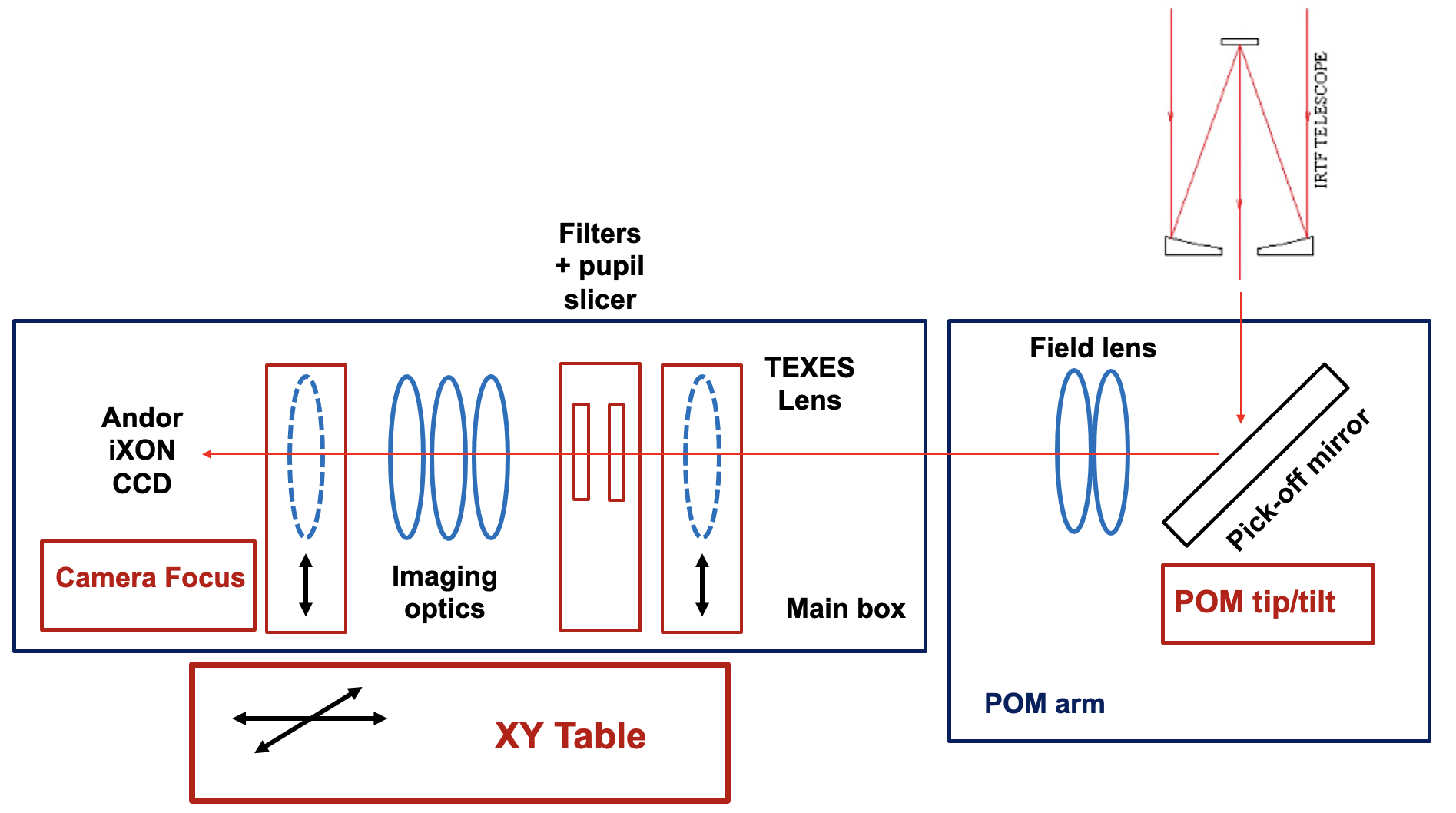}
        \caption{A cartoon of the optical design of FELIX.}
        \label{fig:felix-layout}
    \end{subfigure}

    \begin{subfigure}{\textwidth}
        \centering
        \includegraphics[width=0.7\linewidth]{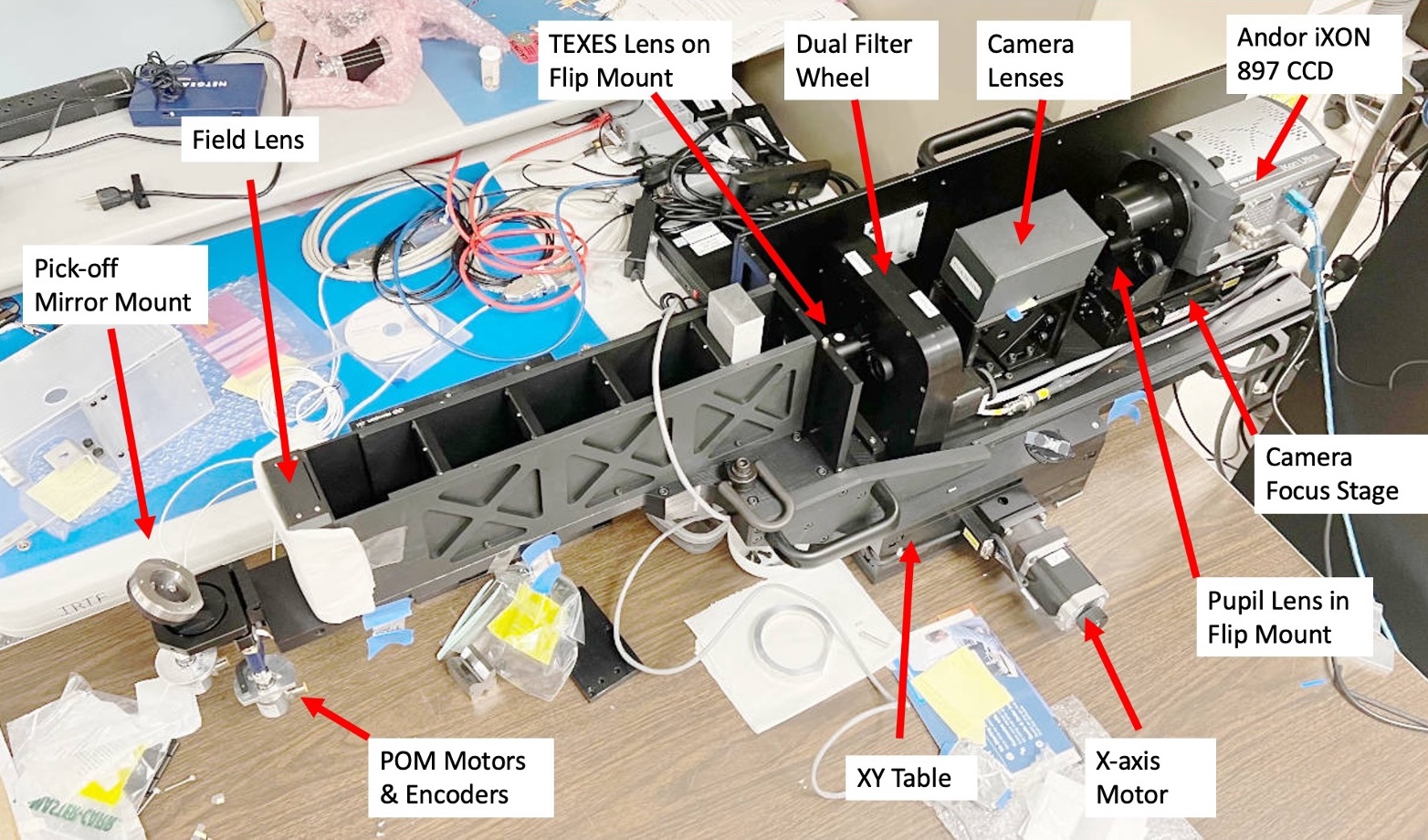}
        \caption{A picture of FELIX in the lab.}
        \label{fig:felix-pic}
    \end{subfigure}
    \caption{FELIX, the facility acquisition camera and low-order WFS. FELIX uses a pick-off mirror to obtain off-axis guide stars. It is converted to a 2$\times$2 SHWFS by placing a pyramid into the pupil plane, which is stored in the filter wheel mechanism. This design allows FELIX to recycle the downstream optics to image the WFS spots and, if desired, to quickly swap the pyramid for one with more facets to obtain higher order correction. FELIX currently only has the 2$\times$2 pyramid.}
    \label{fig:felix}
\end{figure}

FELIX primarily serves as the facility acquisition camera and guider, shown in Fig.~\ref{fig:felix}. It can be converted to a $2\times2$ SHWFS via a mechanism that places a four-sided pyramid into the pupil plane. This allows the WFS mode of FELIX to reuse the downstream imaging optics to form the four Shack-Hartmann spots, and since the pyramid is stored in the filter wheel, it can be swapped in and out easily. The field of view (FOV) of FELIX is \qty{80}{\arcsecond}, which is fed by a pick-off mirror that can either be placed on-axis for target acquisition or moved to acquire an off-axis guide star between \qty{2}{\arcminute} and \qty{4.5}{\arcminute} away. This is suitable for sensing static aberrations in the primary mirror, which are always common path between the WFS and on-axis science instrument, and/or for correcting ground layer turbulence. FELIX has a pixel scale of \qty{0.16}{\arcsecond} and operates at visible wavelengths, with SDSS g', r', i', and z' filters available. We generally use the clear filter for wavefront sensing unless the star is prohibitively bright.

FELIX's instrument software runs on a Linux machine that is mounted on the telescope and reads the camera via a USB camera. The wavefront sensing mode of FELIX was included to automate focus adjustment and to manually monitor static aberrations. As a result, the facility software is designed for running FELIX more like a science imager and so has significant overheads that prevent FELIX from running faster than \qty{1}{\hertz}. FELIX's camera (a 512$\times$512 Andor iXon EM-CCD) is currently capable at running much faster than this, reaching frame rates up to \qty{200}{\hertz} using a subarray. (It is possible to increase the frame rate further using Andor's Camera Link connection, but this requires a dedicated card to capture frames and is not necessary for our use case.) Although the a \qty{1}{\hertz} frame rate is sufficient for active optics, the slope calculations provided by the facility instrument software are also designed for focus control and do not necessarily provide a good estimate of other Zernike modes, especially if the seeing is mediocre.

We installed the Python package pyRTC\footnote{\url{https://github.com/jacotay7/pyRTC}} on FELIX's computer as the real-time control software to close the loop with FELIX. pyRTC was selected for ease of use, and because FELIX is not intended to run at high speeds, the reduced speed from using a high level language like Python is not noticeable. Using pyRTC required adding software interfaces for the camera and ASM along with a modification to the slope calculation process to deal with the fact that the spots in FELIX are tilted about \qty{50}{\degree} with respect to the detector's axes. We also added the ability to record telemetry in the format used for the `imaka GLAO testbed \cite{chun2018sky} along with a few extra functions to assist in chopping the secondary mirror. Preparing pyRTC for use on-sky with FELIX took about one month, most of which was spent debugging the hardware interfaces and testing the observing process with a simulation mode. We were unable to test FELIX and the ASM together in the lab as the former was already installed at the summit and required for regular operations.


\section{On-sky performance} \label{sec:performance}
At their core, active optics and enhanced seeing mode (ESM) are implemented the same way with the only difference being how fast FELIX's camera is running. If a bright off-axis is available, it is possible to begin correcting atmospheric turbulence to enhance the seeing. This is possible even though the guide star is several arcminutes away because most of the turbulence in Maunakea is confined to a thin layer near the ground, so applying a correction with the ASM--which is optically conjugate to near the ground--will improve the image quality over many arcminutes \cite{chun2009mauna}.

\begin{figure}[h]
    \centering
    \begin{subfigure}[t]{0.49\textwidth}
        \centering
        \includegraphics[width=0.99\linewidth]{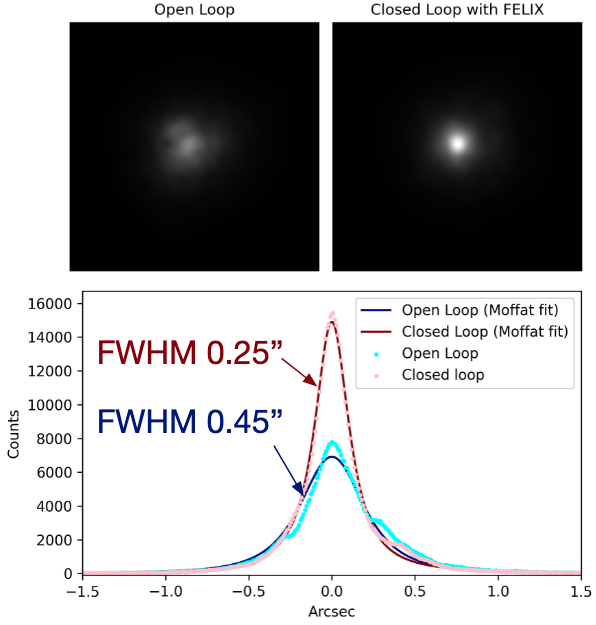}
        \caption{No static aberrations}
        \label{fig:felix-openclose-good}
    \end{subfigure}%
    ~ 
    \begin{subfigure}[t]{0.49\textwidth}
        \centering
        \includegraphics[width=0.99\linewidth]{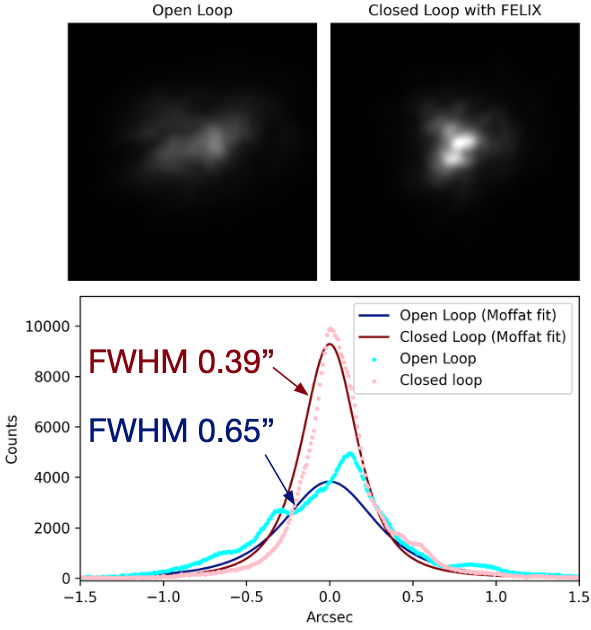}
        \caption{With static aberrations}
        \label{fig:felix-openclose-static}
    \end{subfigure}
    \caption{Images of $\epsilon$ Pegasi (Enif) in open and closed loop with FELIX. The off-axis WFS star in FELIX is about \qty{3}{\arcminute} away. FELIX is running at \qty{97}{\hertz} and is correcting for the first 7 Noll Zernike modes through coma. The seeing was \qty{0.6}{\arcsecond} at \qty{0.5}{\micro\meter} according to MASS/DIMM measurements. The images were taken by stacking \qty{40}{\second} of short exposure (\qty{120}{\milli\second}) frames. In open loop, telescope guiding was enabled, which was updated approximately every \qty{2}{\second}. a) Closing the loop with FELIX improved the FWHM of the PSF by a factor of 1.8, and the equivalent-noise area (ENA) by a factor of 1.5. b) There is a significant amount of structure in the open loop image. The ASM was using a flat command that was measured 0.35 airmasses to the east, so the static corrections are likely compensating for aberrations arising from the primary mirror support structure. Closing the loop with FELIX improves the static aberrations but does not remove trefoil, which is not sensed by the 2$\times$2 spots. The ENA was improved by a factor of 2 in this case.}
    \label{fig:felix-openclose}
\end{figure}

Figure~\ref{fig:felix-openclose} shows open and closed loop images with FELIX. The reference slopes in FELIX change depending on where the off-axis WFS star is located as a result of field curvature in the IRTF focal plane. We discuss this further in Sec.~\ref{sec:challenges}. Consequently, it was necessary to first close the loop with the 12$\times$12 WFS to measure the reference slopes with FELIX. Closing the loop with FELIX at \qty{97}{\hertz} was able to improve the FWHM of the PSF by a factor of 1.8 in \qty{0.5}{\arcsecond} seeing. We were also able to remove most of the static aberrations in the telescope. Trefoil cannot be sensed with the 2$\times$2 pyramid. We are evaluating whether it is worth adding a 3$\times$3 pyramid to FELIX. Since the pyramid is stored in a filter wheel, it would be possible to swap from the 3$\times$3 to the 2$\times$2 if more sensitivity is required. The equivalent-noise area (ENA) was improved by a factors of approximately 1.5 to 2 depending on the atmospheric conditions and static aberrations in the telescope \cite{1983PASP...95..163K}. (Our tests occurred within about 0.4 airmasses from zenith, and we would expect worse static aberrations if the telescope was pointed farther over.) This is consistent with the improvement in throughput that we measure to our facility slit spectrograph, which is shown later in Sec.~\ref{sec:science.glao}.

Our results are similar to the ESM at the Large Binocular Telescope (LBT), which is able to improve the FWHM of the PSF by a factor of two in mediocre 1.5-\qty{2}{\arcsecond} seeing conditions \cite{miller2024enhanced}. LBT's ESM has similarities to our system, using a WFS with a \qty{2}{\arcminute} by \qty{3}{\arcminute} patrol field to sense the first 11 Zernike modes through spherical. LBT also uses ASMs to enhance the seeing but uses a pyramid WFS rather than a SHWFS. We do not know what level of correction we be able to provide in mediocre seeing as the conditions were good throughout the 6 nights of engineering with FELIX and the ASM.

We corrected the first seven Zernike modes and closed the loop with a leaky integrator. Modal control is necessary to avoid injecting high order shapes that cannot be sensed by FELIX. The conversion between actuator commands and Zernike coefficients was measured in the lab by injecting slope offsets, closing the loop with the 12$\times$12 WFS, and recording the resulting commands. The loop rate was determined by the frame rate of the FELIX camera and was thus limited by the brightness of available off-axis guide stars.

For the majority of our run, we used a push-pull interaction matrix that was measured during a period of exceptionally good seeing. In less optimal conditions, we were also able to measure an equivalent interaction matrix and close the loop by using the DO-CRIME method \cite{lai2021crime}. There was no noticeable difference in image quality delivered by the push-pull versus DO-CRIME interaction matrix. Ultimately, we plan to use a theoretical interaction matrix as the registration of the ASM is quite forgiving on the large subapertures within FELIX. Differences in the orientation of the ASM can be measured with DO-CRIME or a different method (for example, CACOFONI \cite{zhang2025cacofoni}) and used to scale and rotate the theoretical interaction matrix.

DO-CRIME is a method of extracting the interaction matrix through turbulence by injecting white noise into the actuators. Although it can be used as-is with FELIX running at \qty{100}{\hertz}, it must be used with caution at higher loop rates due to the slow open-loop response of this generation of TNO HVR actuators. The actuators take \qty{10}{\milli\second} to settle in open loop \cite{zhang2024lab}. With the 12$\times$12 WFS running at \qty{1}{\kilo\hertz}, we have been able to generate "lagging" DO-CRIME to interaction matrices by waiting a few milliseconds for the actuators to settle before recording the response. Methods for generating interaction matrices with the HVR actuators in the context of using larger ASMs are elaborated in Ref.~\cite{zhang2025cacofoni}.

\subsection{Chopping the ASM} \label{sec:chop}
IRTF regularly observes in the mid-infrared with its 2-\qty{20}{\micro\meter} imager and low-resolution spectrograph MIRSI, which is primarily designed to determine the sizes and albedos of near-earth objects as part of NASA's planetary defense program \cite{hora2024design}. At these wavelengths, it is beneficial to quickly measure the sky background by chopping the secondary mirror several times per second \cite{hoffmann1994astronomical}.

\begin{figure}[h]
    \centering
    \begin{subfigure}[t]{0.49\textwidth}
        \centering
        \includegraphics[width=0.7\linewidth]{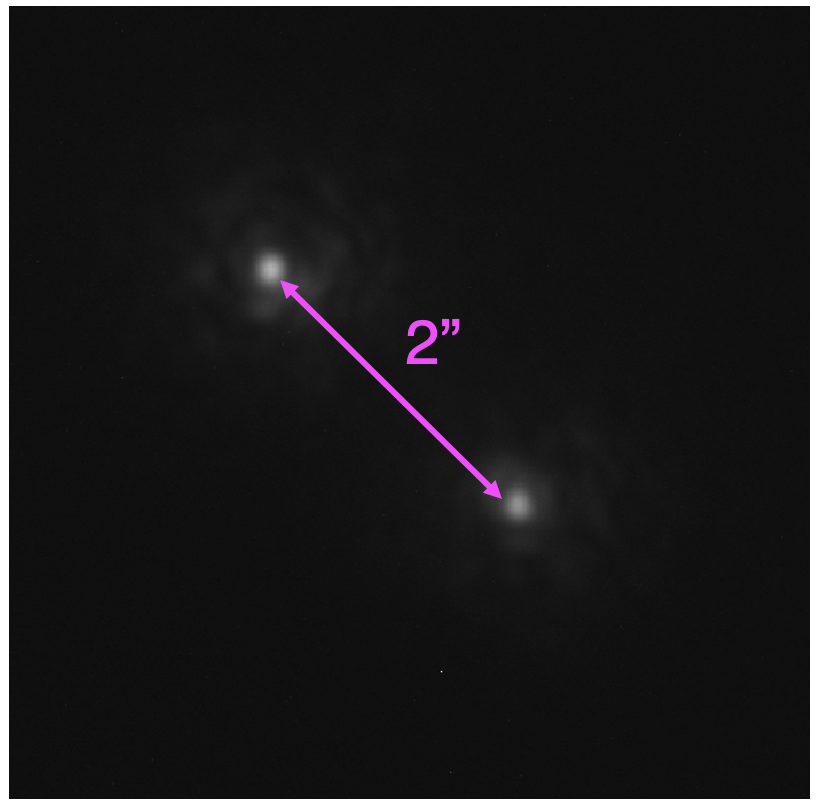}
        \caption{Chopping with the 12$\times$12 WFS}
        \label{fig:chop-img-12x12}
    \end{subfigure}%
    ~ 
    \begin{subfigure}[t]{0.49\textwidth}
        \centering
        \includegraphics[width=0.7\linewidth]{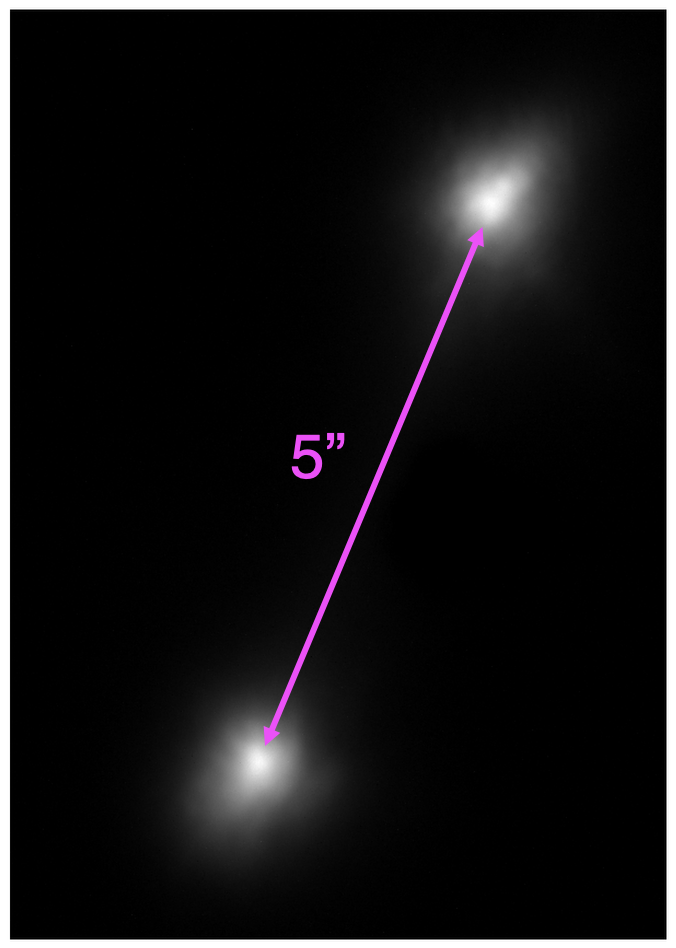}
        \caption{Chopping with FELIX}
        \label{fig:chop-img-felix}
    \end{subfigure}
    \caption{Long exposure focal plane images ($\lambda=$\qty{1.65}{\micro\meter}, $t \geq$ \qty{10}{\second}) when chopping the ASM at \qty{4}{\hertz} in closed loop. a) The chop is performed by changing the slope offsets in the RTC to translate the spots in the subapertures of the 12$\times$12 WFS while the AO loop is closed at \qty{1}{\kilo\hertz}. The image is diffraction-limited on both ends of the chop. b) Because the loop rate of FELIX is slower and the chop amplitude is very large, chopping with FELIX is performed by throwing the ASM in open loop and blindly changing the region of interest to close the loop at each end of the chop. The \qty{5}{\arcsecond} chop amplitude uses about 75\% of the range of actuator stroke. A slight overshoot can be seen on both ends of the chop because the images were stacked blindly without rejecting frames with poor image quality. }
    \label{fig:chop-imgs}
\end{figure}

Figure~\ref{fig:chop-imgs} shows long exposure images taken while chopping the ASM at \qty{4}{\hertz}. We began to test the ASM's ability to chop by closing the loop with the 12$\times$12 WFS and updating the reference slopes to inject tilt following a square wave at \qty{4}{\hertz} (Fig.~\ref{fig:chop-img-12x12}). The ASM was thus in closed loop through the ramp of the chop. Using this method, the amplitude was limited by the size of the subapertures in the WFS, which are \qty{5}{\arcsecond} across, rather than actuator stroke. Because of the large stroke of the actuators and the lack of a field stop, we avoided pushing the spots too close to the edges of the subapertures to prevent the ASM from accidentally driving the spots very far away from their intended positions.

A new chopping mount is currently being developed for the static secondary mirror, but chopping with the ASM would allow us to observe with MIRSI without a top end exchange (for point sources). There are no strict requirements for chopping. However, an amplitude of 3-\qty{4}{\arcsecond} is desirable to separate an AB pair produced by a point source. The necessary chop frequency is less clear and is being investigated. A minimum frequency of \qty{1}{\hertz} and ideally 5-\qty{10}{\hertz} would be desirable for measuring the sky background at mid-infrared wavelengths (Ref.~\cite{hoffmann1994astronomical} and a private communication with John Rayner).

After FELIX was prepared for use with the ASM, we tested an increased chop amplitude. We threw the ASM in open loop and then closed the loop on the ends of the chop by changing the region of the detector used to define the subaperture masks. This process allowed us to reach higher amplitudes compared to remaining in closed loop using the current control servo; in closed loop with FELIX, it is harder to keep up with the ramp of the chop compared to using the 12$\times$12 WFS due to the lower frame rate. The resulting image is shown in Fig.~\ref{fig:chop-img-felix}. Note that the wavefront error requirements are significantly relaxed at \qty{10}{\micro\meter}. The star appears to be slightly smeared in Fig.\ref{fig:chop-img-felix}, which indicates an overshoot in the open loop chop. In the future, the duty cycle may be increased by using an improved control servo with FELIX and remaining in closed loop throughout the entire chop, much like we have done for the $12\times12$ WFS.


\section{First science with the ASM} \label{sec:science}

\subsection{Static mode} \label{sec:science.static}

\begin{figure}[h]
    \centering
    \includegraphics[width=0.75\linewidth]{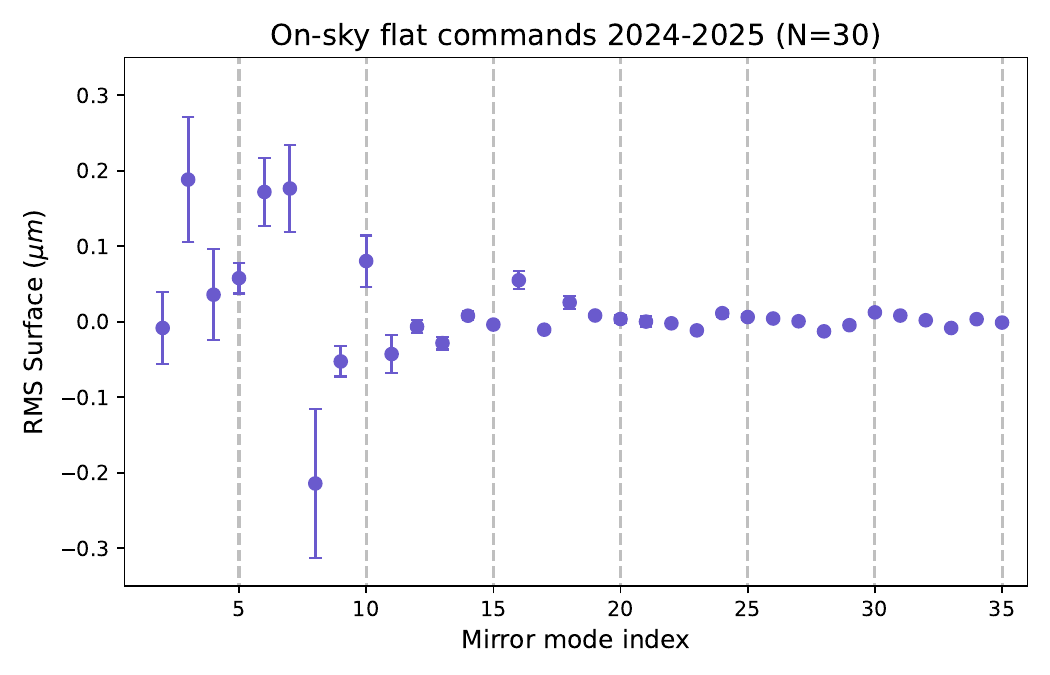}
    \caption{On-sky flat commands of IRTF-ASM-1 over the past year across many telescope pointings (up to 1.5 airmasses in different cardinal directions). The flat commands are measured by closing the loop with the 12$\times$12 WFS and averaging over at least \qty{10}{\second} of data. The commands are plotted in terms of mirror mode coefficients which are ordered in increasing power at higher spatial frequencies (see Fig. 7 of Ref.~\cite{lee2024first}). Indexing of the mirror modes starts at zero, but the first two modes are omitted from the plot as they mostly correspond to tip-tilt and are thus much larger than other terms. The high order shape of the ASM is very stable.}
    \label{fig:flat-cmds}
\end{figure}

FELIX is designed to have complete sky coverage in the sense that an off-axis guide star should always be accessible in its patrol field. It is capable of sensing a  $V\sim18$ star with \qty{10}{\second} of exposure time. Once the ASM is fully commissioned to work with FELIX, active optics should always be available at a minimum. It is still useful to be able to simulate a static secondary mirror with ASM for rare instances in which FELIX is unavailable, which we call ``static mode." We have been monitoring the on-sky flat commands over the past year to understand whether the static shape of the ASM is stable. Figure~\ref{fig:flat-cmds} shows that the high order shape has been very stable over a year and across many telescope pointings. Most of the variance is in the first 10 modes, which roughly correspond to low-order Zernike modes through trefoil, which is likely from the ASM compensating for static aberrations in the primary mirror. Most of these modes are sensed with FELIX, although a trefoil would require a 3$\times$3 pyramid to be inserted. 

A Target of Opportunity (ToO) observation was triggered during our engineering time to observe a stellar occultation. Due to the time-critical nature of the observation and the fact that active correction with the ASM and FELIX had not been thoroughly tested, we operated the ASM in static mode. Half an hour prior to the observation, we pointed to a bright star near where the event would take place and took a new set of flat commands by closing the loop with the 12$\times$12 WFS and averaging over \qty{10}{\second} of data. The star was about 1.1 airmasses southeast of zenith. Then, the telescope was pointed to zenith to perform an instrument change to MORIS,\footnote{MORIS is mounted on the side of SpeX, a spectrograph that is described in Sec.~\ref{sec:science.glao}.} a high speed visible light imager designed to observe occultation and transit events \cite{gulbis2011first}. Including overheads for target acquisition, it took \qty{10}{\minute} to prepare for the instrument change and ToO program. The observations lasted about an hour during which the image quality appeared to be stable. We were able to successfully observe the occultation event.

\subsection{Enhanced seeing mode} \label{sec:science.glao}
As we demonstrated previously, FELIX and the ASM are capable of improving the seeing by correcting turbulence in the ground layer. To quantify how this translated to improved throughput in IRTF's seeing-limited instruments, we swapped out the 12$\times$12 WFS and used SpeX, a medium resolution near-infrared slit spectrograph and the most popular instrument at IRTF \cite{rayner2003spex}. Our observations were performed with the \qty{0.3}{\arcsecond} slit in SXD mode which provides a resolving power of 2,000 from 0.70-\qty{2.55}{\micro\meter}. We measured spectra in open and closed loop, and then reduced the data using the Spextool IDL package \cite{cushing2004spextool} . The spectra are not telluric corrected as we did not observe an A0V star. We pointed to the Pleiades and used Alcyone as the off-axis guide star, placing a nearby A-type star on the slit of SpeX.

\begin{figure}[h]
    \centering
    \includegraphics[width=0.75\linewidth]{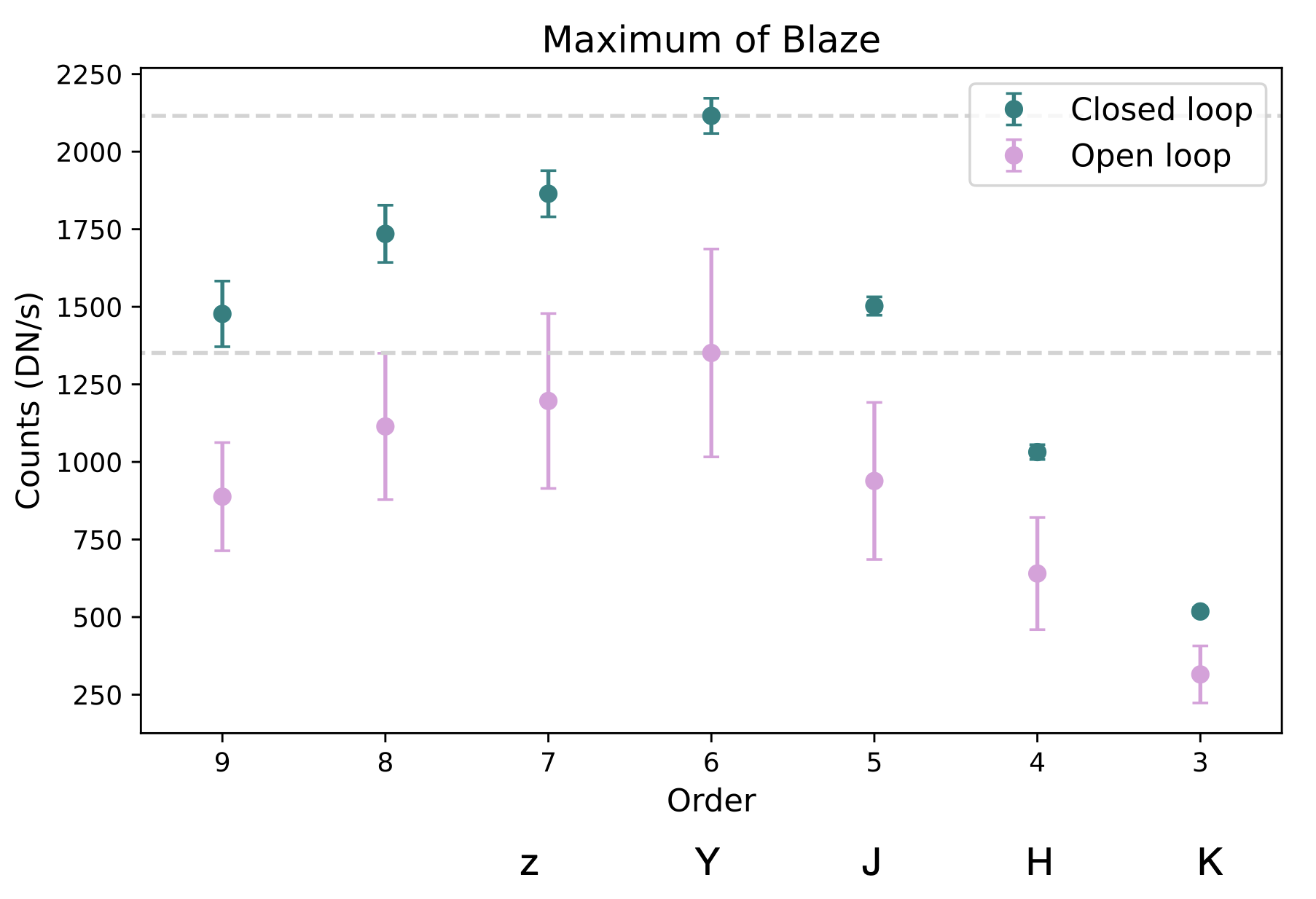}
    \caption{Increase in throughput in the near-infrared cross-dispersed facility spectrograph SpeX. We closed the loop with FELIX at \qty{180}{\hertz} on an off-axis guide star that was about \qty{2.5}{\arcminute} away. Five spectra were taken each in open and closed loop. The spectra were fully reduced, summing over the flux in the \qty{0.3}{\arcsecond} slit, and the maximum number of counts in each of the spectral orders was measured. The closed loop image quality was much more stable than in open loop and the extracted number of counts was improved by a factor of 1.6. Also note that a new set of flat commands was taken shortly before the observation so the open loop image quality does not include static aberrations that are typically found in the IRTF primary mirror. Correcting for static aberrations would improve the throughput further as shown in Fig.~\ref{fig:felix-openclose}.}
    \label{fig:spex-throughput}
\end{figure}

Figure~\ref{fig:spex-throughput} show that the throughput of SpeX is improved by a factor of 1.6 correcting ground layer turbulence with the ASM. These results are not a fair comparison to IRTF's image quality with a static secondary mirror because we took a new set of flat commands shortly before these tests. The new flat commands eliminated static aberrations that are usually present in the primary mirror. We cannot directly measure the improvement SpeX's throughput from removing static aberrations because it is difficult to disentangle low-order aberrations from the telescope, the ASM, and differences in alignment. However, as shown in Sec.~\ref{sec:performance} and in Ref.~\cite{dinh2024measuring}, these aberrations can degrade the FWHM of the image by a factor of two or more so the total gain of the ASM over the fixed secondary is probably closer to a factor of three to four. 

\section{Challenges} \label{sec:challenges}
\begin{figure}[h]
    \centering
    \begin{subfigure}{0.49\textwidth}
        \centering
        \includegraphics[width=0.99\linewidth]{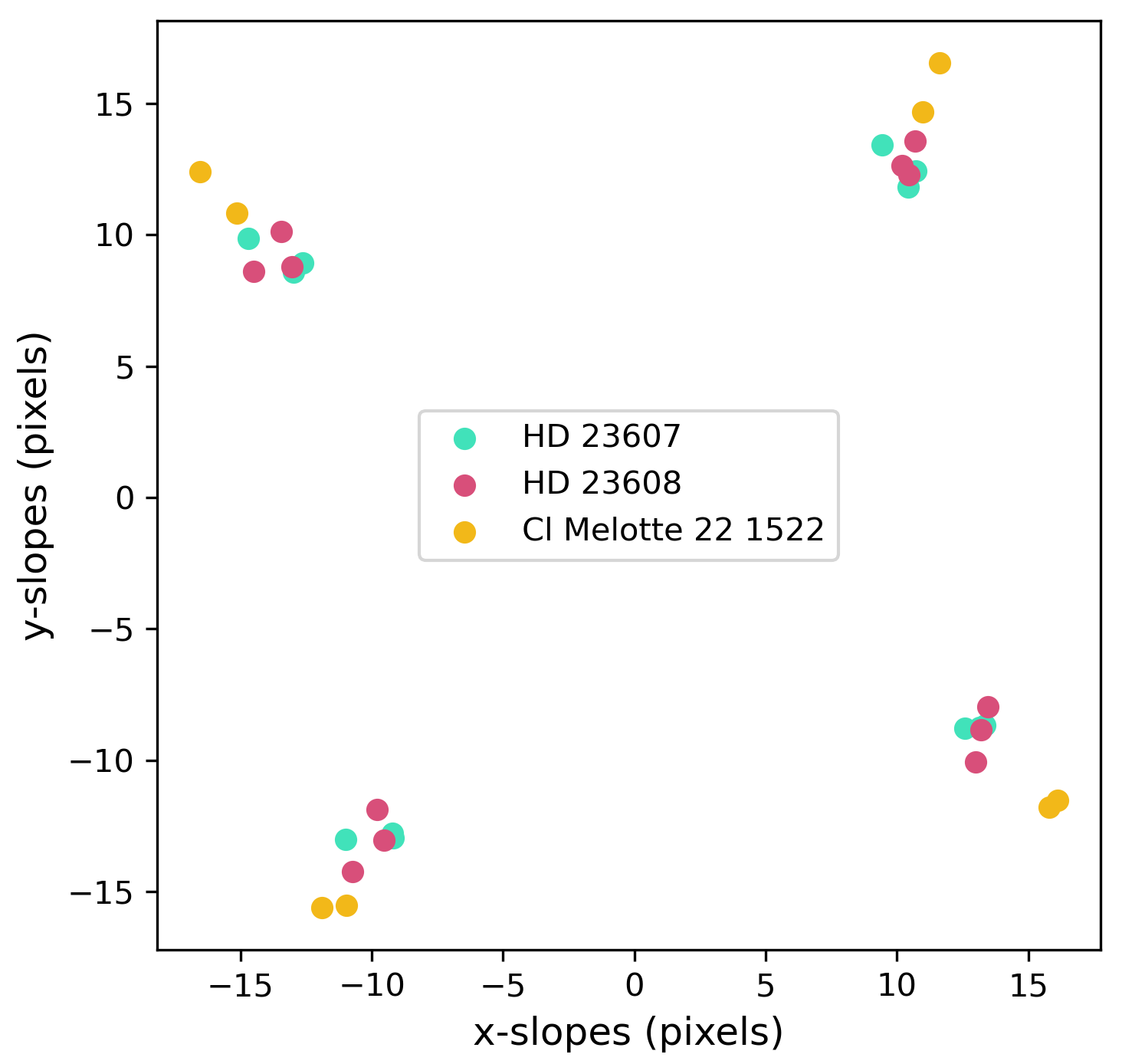}
        \caption{Reference slopes for different off-axis stars}
        \label{fig:refslopes-plot}
    \end{subfigure}%
    ~ 
    \begin{subfigure}{0.49\textwidth}
        \centering
        \includegraphics[width=0.99\linewidth]{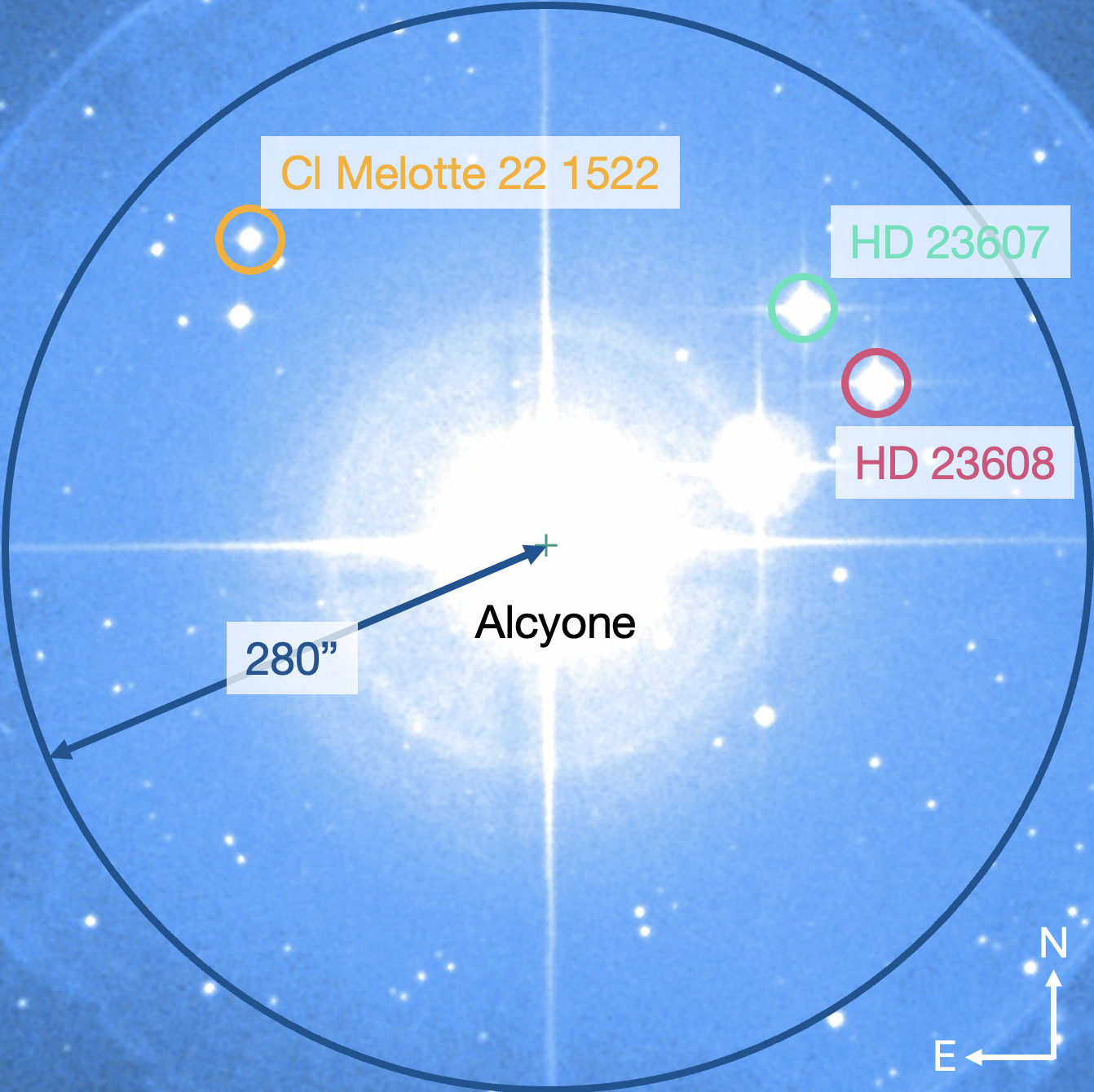}
        \caption{Off-axis stars in the FELIX's patrol field }
        \label{fig:refslopes-stars}
    \end{subfigure}
    \caption{Reference slopes in FELIX for different off-axis WFS star positions. The coordinates are relative to the center of the region of interest containing all of the subapertures. a) The reference slopes for each of the four spots were measured by closing the loop with the 12$\times$12 WFS and averaging the slopes computed in FELIX over 7-\qty{10}{\second}. The reference points for each star (different colors) were measured across two to three separate nights hence multiple sets of four points per star. Roughly speaking, the two stars that are closer together in the sky have similar reference slopes whereas the third star requires more focus. b) Positions of the three off-axis WFS stars on the sky. The star $\eta$ Tauri (Alcyone) was used on-axis with the 12$\times$12 WFS to produce to reference slopes in FELIX. FELIX's \qty{280}{\arcsecond} radius patrol field is indicated on the image. The image is from the Digitized Sky Survey (DSS).}
    \label{fig:refslopes}
\end{figure}

\begin{figure}
    \centering
    \includegraphics[width=0.7\linewidth]{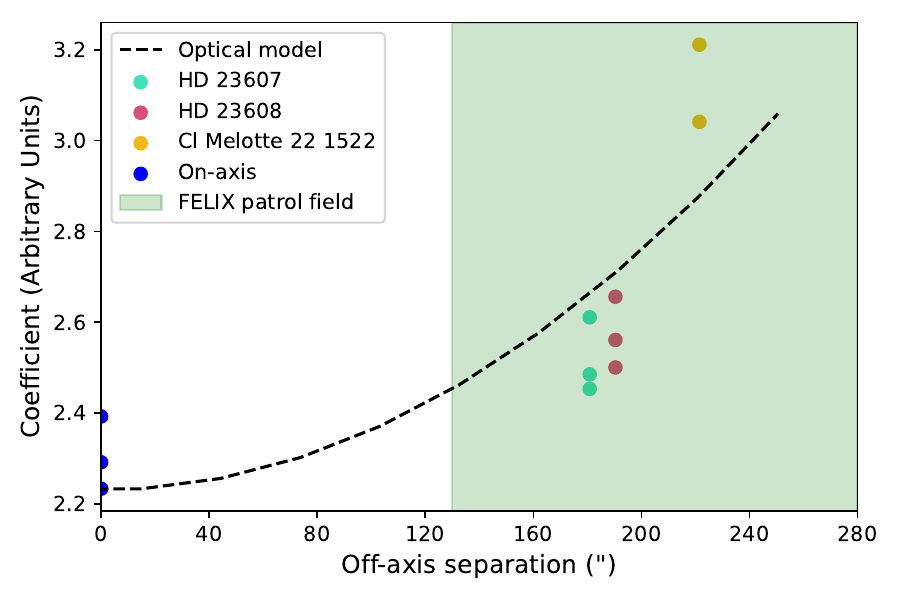}
    \caption{Focus coefficient in the reference slopes compared to the Zemax model of the telescope's field curvature. The focus is offset from zero at zero separation because of the relative focus between FELIX and the 12$\times$12 WFS. Although there is very rough agreement, there appears to be a component that is not explained by the field curvature.}
    \label{fig:focus-model}
\end{figure}

The biggest problem we have faced is obtaining a good set of reference slopes for FELIX. While we have the 12$\times$12 WFS, we are able to obtain a reference by closing the loop on-axis and measuring the slopes in FELIX. This cannot be done when a science instrument is on-axis. We have noticed that the reference slopes change depending on which off-axis star is being used. Figure~\ref{fig:refslopes} shows the reference slopes for different off-axis guide stars in the Pleiades. Most of the change appears to be in focus, which is not surprising given that there is field curvature in IRTF (Fig.~\ref{fig:focus-model}). The field curvature can be calibrated out by a lookup table, but there appear to be additional aberrations that are not explained by the optical model of the telescope. There is also significant spread in the measured reference slopes even on the same target.

When the reference slopes were not set correctly, closing the loop with FELIX made the image worse than the seeing. We were limited in how much we could assess the improvement to instrument throughput because we had to manually calibrate the reference slopes by bringing the star out of focus in the slit viewing cameras. We require a method to verify that FELIX is driving the spots to the correct position based on images in the slit viewer, which is challenging because the slit viewing cameras have a pixel scale of \qty{0.1}{\arcsecond}/px to sample seeing-limited PSFs. It is difficult to discern structure in the PSF in instances when FELIX is able to provide 0.2-\qty{0.3}{\arcsecond} image quality. Furthermore, the readout of the slit viewing cameras is also slow, which limits the frame rate of our images to perhaps \qty{1}{\hertz}. The process of tuning the reference slopes must also be fast to not be obtrusive to science observations. We are considering different methods of wavefront sensing using the slit viewing cameras, for example, curvature wavefront sensing by chopping the ASM and applying focus at each end of the chop.

Additionally, IRTF does not have existing AO capability and thus cannot to run an AO system in the traditional sense, where an operator with AO expertise is present to care for the system. Active optics and ESM may essentially be implemented the same way, but automating the latter requires more caution in tuning the loop. The parameters of the control servo along with the WFS frame rate and binning need to be adjusted depending on the brightness of the off-axis guide star and seeing. These parameters can be pre-calculated based on simulations with minor adjustments by the telescope operator if necessary.

GLAO is best implemented with multiple wavefront sensors to disentangle turbulence in the ground layer from that in the upper atmosphere. Our method of enhancing the seeing relies on most of the turbulence being near the ground so that blindly applying correction sensed by the off-axis FELIX to the ASM can still improve the on-axis image. In cases when the turbulence is much stronger in the upper atmosphere, e.g., when the jet stream is overhead, attempting to do this will make the on-axis image worse. We can use measurements from the combined Multi-Aperture Scintillation Sensor (MASS) and Differential Image Motion Monitor (DIMM) at the summit of Maunakea to monitor where the turbulence is strongest on a nightly basis \cite{kornilov2007combined,schock2009thirty}. If the turbulence is strong at high altitude, we can disable ESM with FELIX.

\section{Conclusions and future work} \label{sec:conclusion}
IRTF-ASM-1 has been robust and continues to function consistently well. By running FELIX at 90-\qty{180}{\hertz} on an off-axis star about \qty{3}{\arcminute} away and correcting the first 7 Zernike modes, we are able to improve the throughput of IRTF's slit spectrograph SpeX by a factor of two in enhanced seeing mode. The ASM is also capable of removing static aberrations in the primary mirror to improve the FWHM of the image by another factor of two or more. We performed time-critical science observations with the ASM in static mode and have observed that the high order shape of the ASM is stable over a year and across different telescope pointings (up to 1.5 airmasses).

Active optics will likely be prepared for nightly use around Spring 2027 pending a new hexapod and further work to streamline the system so it can be used by a telescope operator. The next engineering run with the ASM will be in Spring 2026, during which we plan to test improved calibrations for FELIX's reference slopes and our ability to run active optics mode with reduced human input. Complete facility integration of ESM would require at least two more years of effort if it is supported by IRTF. We also mention that the University of Hawai'i's 88-inch telescope (UH88) will be obtaining its ASM shortly. The UH88 ASM completed integration and lab acceptance testing at TNO and will be delivered in early 2026 following end-to-end testing with its new electronics.

\acknowledgments 

This work and two of its authors (Lee and Zhang) have been supported by a National Science Foundation Advanced Technology and Instrumentation award (NSF-1910552). The Infrared Telescope Facility is operated by the University of Hawaii under contract 80HQTR19D0030 with the National Aeronautics and Space Administration. The first author of this manuscript is currently funded by the NASA Infrared Telescope Facility. We also wish to recognize and acknowledge the very significant cultural role and reverence that the summit of Maunakea has always had within the indigenous Hawaiian community. The opportunity to conduct observations from this mountain is an enormous privilege, and we feel gratitude for our ability to study astronomy from Maunakea.

\printbibliography

\end{document}